\documentclass[12pt]{article}
\usepackage[utf8]{inputenc}

\usepackage{hyperref}
\usepackage[super,sort,compress]{cite}
\usepackage{amssymb}
\usepackage{xcolor}
\usepackage{amssymb}
\usepackage[utf8]{inputenc} % set input encoding (not needed with XeLaTeX)
\usepackage{graphicx} % support the \includegraphics command and options
\usepackage{algorithm}
\usepackage{algorithmicx}
\usepackage{algpseudocode}
\usepackage{amsmath}
\usepackage{booktabs} % for much better looking tables
\usepackage{array} % for better arrays (eg matrices) in maths
\usepackage{paralist} % very flexible & customisable lists (eg. enumerate/itemize, etc.)
\usepackage{verbatim} 
\usepackage{float}
\usepackage{placeins}
\usepackage{underscore}           % Only needed if you use pdflatex.
\usepackage[all,knot,poly]{xy}
\usepackage[english]{babel}
\usepackage[T1]{fontenc}
\usepackage{graphicx}
\usepackage{epstopdf}
\usepackage{subfigure}
\usepackage{doi}
\usepackage{appendix}
\usepackage{pdfpages}
\usepackage{placeins}
\usepackage{graphicx}
\usepackage{verbatim}
\usepackage{mathtools}
\usepackage{amssymb, amsmath}
\usepackage{tikz}
\usepackage{xfrac}
\usetikzlibrary{decorations.pathreplacing, shadows, shapes}
\usepackage{xfrac}
\usetikzlibrary{decorations.pathreplacing, shadows, shapes}
\usepackage[qm]{qcircuit}
\usepackage{comment}
\usetikzlibrary{shapes,snakes}
\xyoption{arc}
\usepackage[qm]{qcircuit}

\usepackage[utf8]{inputenc}
\usepackage{pgfplots}

\usepackage{algorithm}
\usepackage{algpseudocode}

\newcommand{\test}{\text{test}}
\newcommand{\happy}{\text{happy}}
\newcommand{\sad}{\text{sad}}
\newcommand{\fun}[1]{\textsc{#1}}
\newcommand{\G}{\mathcal{G}}
\newcommand{\return}{\textbf{return} }
\newcommand{\R}{\mathbb{R}}

\def\be{\begin{eqnarray}}
\def\ee{\end{eqnarray}}

\newcommand{\ket}[1]{|#1\rangle}

\title{Facial Expression Recognition on a  Quantum Computer}
\author{Riccardo Mengoni, Massimiliano Incudini, Alessandra Di Pierro}
\date{\small{Department of Computer Science\\ University of Verona, Italy}}

\begin{document}
	
%	\begin{titlepage}
		\maketitle
%	\end{titlepage}

%
%\begin{document}

%\title{Facial Expression Recognition on a  Quantum Computer
%%\thanks{Grants or other notes
%%about the article that should go on the front page should be
%%placed here. General acknowledgments should be placed at the end of the article.}
%}
%%\subtitle{Do you have a subtitle?\\ If so, write it here}
%
%%\titlerunning{Short form of title}        % if too long for running head
%
%\author{Riccardo Mengoni        
%\and Massimiliano Incudini 
%\and Alessandra Di Pierro 
%}


%\institute{R. Mengoni \at
%          Department of Computer Science, University of Verona,
%          Verona, Italy\\
%              \email{riccardo.mengoni@univr.it}           %  \\
%%             \emph{Present address:} of F. Author  %  if needed
%           M. Incudini \at
%             Department of Computer Science, University of Verona, 
%             Verona, Italy\\
%             \email{massimiliano.incudini@studenti.univr.it}                
%             \and
%             A. Di Pierro \at
%             Department of Computer Science, University of Verona, 
%             Verona, Italy\\
%             \email{alessandra.dipierro@univr.it}    
%}
%
%\date{}

%\maketitle

\begin{abstract}
We address the problem of facial expression recognition  and show a possible solution using a quantum machine learning   approach.
%We apply a method introduced in [\citen{Fingerhuth}] 
In order to define an efficient classifier for a given dataset, our approach substantially exploits quantum interference. By representing face expressions via graphs, we define a classifier as a quantum circuit that manipulates the graphs adjacency matrices encoded into the amplitudes of some appropriately defined quantum states.

We discuss the accuracy of the  quantum classifier  evaluated on the quantum simulator available on the IBM Quantum Experience cloud platform, and compare it with the accuracy of one of the best classical classifier. 

{\bf Keywords}:Quantum Machine Learning, Quantum Computing, Graph Theory, Facial Expression Recognition

\end{abstract}

\section{Introduction}
\label{intro} 
A modern approach to Pattern Recognition is the use of Machine Learning (ML) and, in particular, of supervised learning algorithms for pattern classification. This task essentially consists in assigning a class in a given partition of a dataset to an input value, on the basis of a set of training data whose classes are known.
As witnessed by the emerging of the field of Quantum Machine Learning (QML) 
[\citen{wittek,Schuld-ML,Biamonte17}], Quantum Computation [\citen{ChuangNielsen}] offers a number of algorithmic techniques that can be advantageously applied for reducing the complexity of classical learning algorithms. This possibility has been variously explored for pattern classification, see e.g.  [\citen{Schuld2014}] and the references therein, and the more recent paper [\citen{Park_2020}].

One of the most important application of pattern classification is Facial Expression Recognition [\citen{FER}].
In this field, graph theory [\citen{Bondy:1976:GTA:1097029}] provides a suitable mathematical model of a human face.  
%Graphs  are discrete objects that allow us to schematize a large variety of problems and processes spanning  among many research areas 
In this paper we address the problem of graph classification for implementing the final stage of a facial expression recognition system, namely the stage where an expression, described by a set of features, is assigned to one of several classes representing basic emotions such as anger, happiness, sadness, joy, etc.

Starting from a set of features that were retrieved from a face region in a previous stage, we associate a graph representation to each facial expression by following two alternative strategies: one generates complete graphs, while the other uses a triangulation algorithm to output meshed graphs.
We then define a quantum supervised  learning algorithm that  recognizes input images by assigning a specific class label to each of them.
%In this endeavour we  use some techniques that have been developed in the new growing field of Quantum Machine Learning [\citen{wittek}].

A crucial passage of our  method is the  representation of graphs as quantum states, for which we will make use of the \textit{amplitude encoding} technique, i.e. the encoding of the input features into the amplitudes of a quantum state and their manipulation  through quantum gates. Since a state of $n$ qubits is described by $2^n$ complex amplitudes, such an encoding automatically produces an exponential compression of the data. Combined with an appropriate use of quantum interference, this technique is at the base of the computational speed-up of many quantum algorithms; for example it is responsible for the exponential speed-up in the performance of all quantum algorithms based on the Quantum Fourier Transform [\citen{ChuangNielsen}]. 
However, in our context, an exponential speedup of the overall algorithm is not to be taken for granted, as the nature of the data may require a computationally expensive initialization of the quantum state.

The quantum circuit we construct is inspired by the work in  [\citen{Fingerhuth}]. This circuit performs a classification similar to the k-nearest neighbors classification algorithm used in classical ML, but exploits quantum operations with no classical counterpart, such as those that realize quantum interference: Hadamard gates are used to interfere the new input with the training inputs in a way that a final measurement of a class qubit identifies  the class of the input. As already pointed out in  [\citen{Fingerhuth}], this approach uses quantum techniques for implementing ML tasks rather than simply translating ML algorithms for making them run on a quantum computer.

 We show the results of an experimental testing of our algorithm that we have performed by using the IBM open-source quantum computing software development framework Qiskit [\citen{Qiskit}]. We compare the results obtained on this quantum simulator with those obtained by using a classical algorithm that also uses distances for classifying the data.
Our experiments show that the accuracy of the quantum classification follows very closely that of the classical classification if we use the meshed strategy for representing the input data. In the complete graph approach we observe, instead, a much better performance of the classical algorithm. We will argue that this can be explained in terms of the preliminary encoding of the data into quantum states and the higher error rate in the implementation of the complete graphs approach.

This paper is structured as follows. In Section~\ref{Section1}, we introduce the dataset employed for  face recognition as well as the preprocessing methodology  that extracts a meaningful graph representation of the data. In Section~\ref{Section2} we explain how to define an encoding of face graphs into quantum states. Section~\ref{Section3} and Section~\ref{algos} are devoted to the construction of the quantum classification circuit, the explanation of the algorithms implementing it, and a discussion of our experimental results. Finally, in Section~\ref{Section4} we draw a conclusion and give directions for possible improvements.

 \section{Dataset and Preprocessing}
 \label{Section1}
 
\begin{figure}[htbp]
%  \vspace{-0.2cm}
 	\centering    \includegraphics[width=0.4\textwidth]{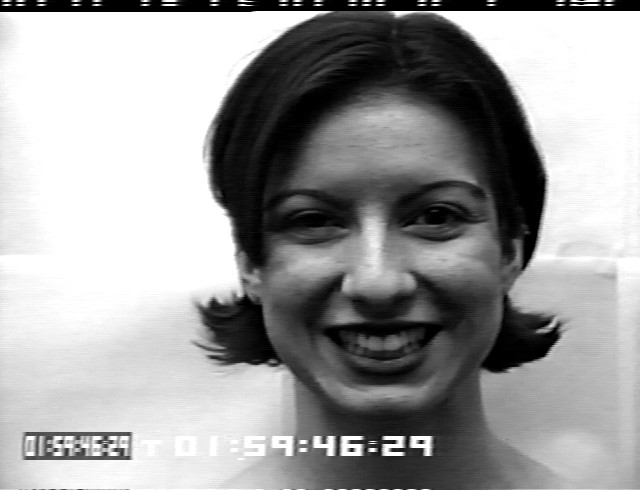}~~~   \includegraphics[width=0.4\textwidth]{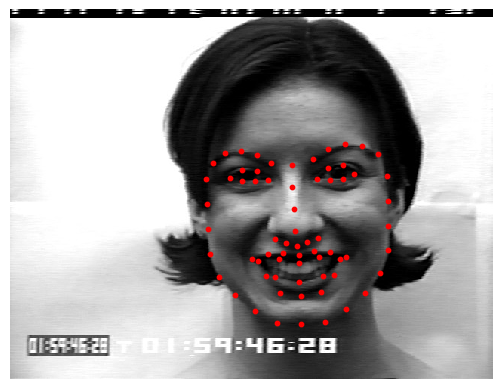}~~~ \caption{An element of the CK+ dataset: happy face (left) and its point cloud (right).}
 	\label{fig:dataset}
 \end{figure}

\begin{figure}[htbp] \vspace{-0.2cm}
 	\centering    \includegraphics[width=0.40\textwidth]{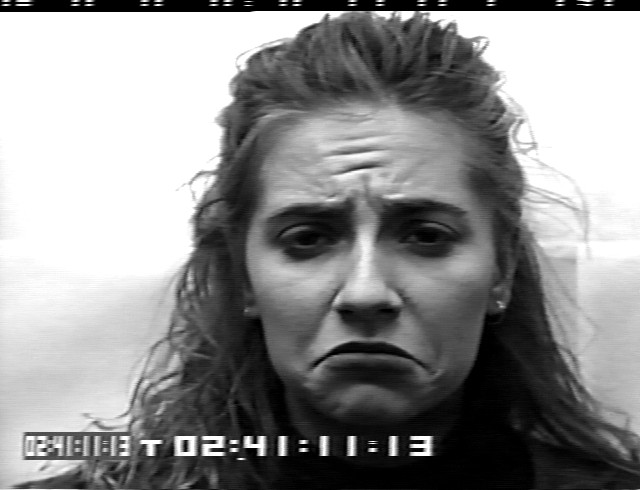}~~~   \includegraphics[width=0.4\textwidth]{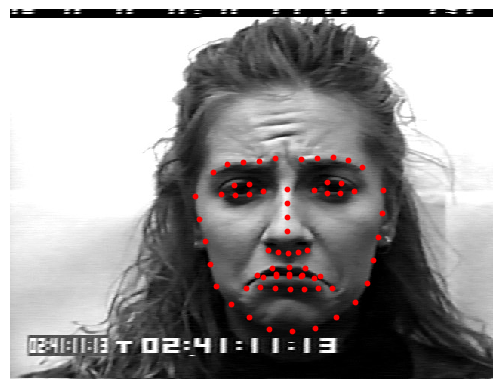}~~~ \caption{An element of the CK+ dataset: sad face (left) and its point cloud (right).}
 	\label{fig:dataset-over}
 \end{figure}

For our experiments we use the freely available  Extended Cohn-Kanade (CK+) database [\citen{CK+}]. This  collects multiple photos of people  labeled by their facial expression, as in the examples  shown in the left-hand side of Fig.~\ref{fig:dataset} and  Fig.~\ref{fig:dataset-over}. 
Each photo is identified with a point cloud  of $68$  points, $(x,y) \in \mathbb{R}^2$, as shown in the right hand side of Fig.~\ref{fig:dataset} and  Fig.~\ref{fig:dataset-over}.

 %\FloatBarrier
 \noindent 
\begin{figure}[htbp]
 	\centering    
 	\includegraphics[width=0.7\textwidth]{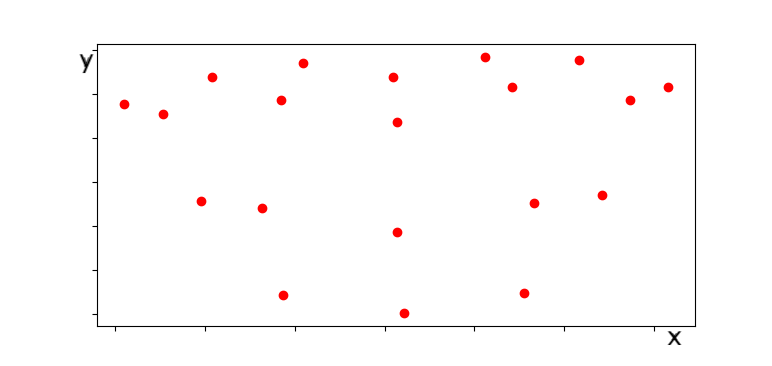}
 	\caption{ Landmark points associated to the happy mouth.  Points are represented in the (x,y) plane.}
 	\label{fig:points}
 \end{figure}

% \FloatBarrier
 \noindent 
 From  this point cloud we select only those 20 points  associated to the mouth\footnote{We will consider here only parts of a facial expression in order to keep the encoding as simple as possible for the sake of the experiments feasibility.}  as shown in Fig.~\ref{fig:points} for the happy expression.
 
 %\FloatBarrier
 \noindent 

With the objective of using this dataset for the inputs of a quantum circuit implementing a classifier of expressions, we first associate a graph to each object of the data set by following two alternative strategies.  
 
The first strategy considers the weighted complete graph whose vertices are the $n$ landmark points of the mouth and whose edge-weights $w_{ij} $ correspond to the Euclidean distance of vertex $i$ from vertex $j$. 
Since a complete graph with $n$ vertices has $n(n-1)/2$ edges, its construction requires $O(n^2)$ steps.
 
 The second strategy is based on the Delaunay triangulation of a set $S$ of $n$ points in $\R^2$ [\citen{Delaunay}]. 
 This is a technique for subdividing a planar object into triangles (and a general geometric object in $\R^d$ into simplices), which constructs a partition of $\R^2$ in triangles (or polyhedral in $\R^d$) as follows: for each point $p$ in the set $S$, consider the convex hulls of the set of points that are closer to $p$ than to any other point in $S$, with respect to the Euclidean distance; then take all the convex hulls together with their faces. The most straightforward algorithm finds the Delaunay triangulation of a set of $ n $ points  in $ O(n^2) $  by randomly  adding one vertex at a time and triangulating  again the affected parts of the graph. However, it is possible to improve this algorithm and reduce the runtime to  $O(n \log n)$ as shown in  [\citen{deBerg}].
By applying the Delaunay triangulation to the points of the mouth of the facial expressions, we obtain meshed weighted graphs, where weights are the same as those used for the complete graphs. 
% The complexity of the triangulation algorithm is $ O(n \log n) $. %%rev2: EXPLAIN!%%%
 
\begin{figure}[htbp] 
	\centering   \textbf{Complete graph}\\ \includegraphics[width=0.7\textwidth]{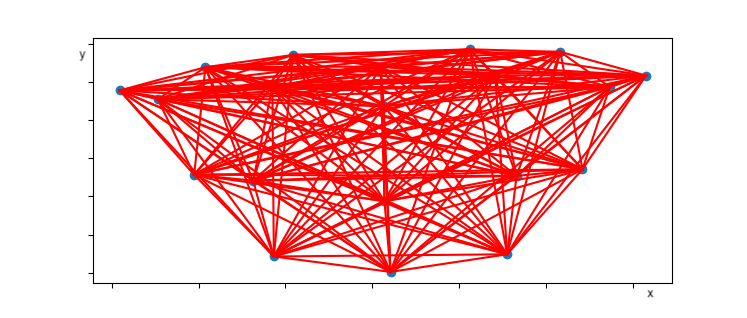}\\ \vspace{.5 cm}
	\textbf{Meshed graph}\\
	\includegraphics[width=0.65\textwidth]{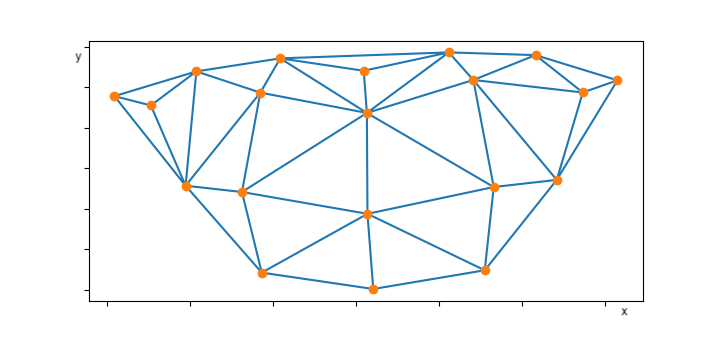}
	\caption{Complete and meshed graphs  obtained from the mouth  landmark points of the happy face, using all the 20 nodes that identify the mouth.}
	\label{fig:graphs}
\end{figure}

Given a mouth landmark point cloud as in Fig.\ref{fig:points}, the outputs of these two strategies are shown  in Figure~\ref{fig:graphs}.

%\FloatBarrier
%\noindent 
 
Clearly, one can expect that a classification based on the complete graphs strategy gives a higher accuracy than the meshed one, as it can exploit a richer description of the data. As we will see later, although this is true for the classical case, the quantum algorithm we present may achieve a better accuracy with the meshed strategy, due to a lower error rate occurring in this case (matrices are sparser than those for complete graphs).

 \section{{Encoding Graphs   into Quantum States}}
 \label{Section2}
 
 Consider an undirected simple  graph $ G = (V_G,E_G) $,  where $ V_G $ is the set of  vertices  and $ E_G $ the set of edges in  $ G $. By fixing an ordering of the vertices $ \{v_i\}_{i=1,..n}$,  a graph  $ G $  is uniquely  identified by its adjacency matrix  $ A_G $ with generic element
 \begin{equation*}
 a_{ij}=
 \begin{cases}
 1 & \text{if } e_{ij}\in E_G,
 \\
 0 & \text{if } e_{ij}\notin E_G.
 \end{cases}
 \end{equation*}
 Therefore if the cardinality  of  $ V_G $ is $ n$, i.e. the graph G has n vertices, then   $ A_G $  is a  n$\times$n square matrix with zeros on the diagonal.
 Since we are dealing with  undirected simple  graphs, $ A_G $ is also symmetric and this means that the meaningful information  about graph $ G $ is contained in the $ d=(n^2-n)/2$ elements of the upper triangular part of $ A_G $.
We can now vectorize those elements  and rename them as  follows:
 \begin{equation}\label{Eq:vectorization}
 \mathbf{a}_G=\left( a_{12}, a_{13}, .., a_{1n}, a_{23},.., a_{2n},.., a_{(n-1)n} \right) ^T=\left( g^{(G)}_1, g^{(G)}_2,.., g^{(G)}_{d} \right)^T
 \end{equation}
 where
 \[
g^{(G)}_{k} \equiv  a_{i,j} \mbox{  for  }  k= i\times n-\frac{i(i+1)}{2} - n + j.
\] 
 
Following the approach in [\citen{QSVM}], from vector $ \mathbf{a}_G $  we construct  a  quantum state $ \ket{G} $ associated to graph $ G $ by encoding the elements of the adjacency vector into the amplitudes of the quantum state:
 \begin{equation}\label{Eq:encoding}
 \ket{G} = \dfrac{1}{\gamma}\sum_{k=1}^{d} g_k \ket{k},
 \end{equation}
 where $ \gamma $ is a normalization constant  given by
 \begin{equation}\label{Eq:normalization}
 \gamma=\sqrt{\sum_{\{k\  \mathrm{s.t}.\  g_k\neq 0\}} |g_k|^2}.
 \end{equation}

\noindent
 This encoding can be extended to the case of weighted graphs $ G = (V_G,E_G,w_{ij})$, where $ w_{ij}\in \mathbb{R}_{\geq 0} $ is the weight associated to the edge $ e_{ij}\in E_G$. 
 In this case, the adjacency matrix is defined as 
 %\begin{equation}
 \[
 a_{ij}=
 \begin{cases}
 w_{ij} & \text{if } e_{ij}\in E_G,
 \\
 0 & \text{if } e_{ij}\notin E_G.
 \end{cases}
 \]
 %\end{equation}
 
Quantum state encoding is then performed as in  Equations~(\ref{Eq:vectorization}), (\ref{Eq:encoding}), (\ref{Eq:normalization}).
 This  allows us to represent a classical vector of $d$ elements into  a  quantum state of $N=\lceil \log(d)\rceil$ qubits, i.e., in our case, with a number of qubits that grows linearly with the number $n$ of the graph vertices.
 
 On the IBM Qiskit framework [\citen{Qiskit}], states expressed by Equation~(\ref{Eq:encoding}) are realized by using a method proposed by Shende et al.  in [\citen{Shende}].  This method is based on an asymptotically-optimal algorithm for the initialization of a quantum register, which exploits the fact that  an arbitrary $n$-qubit state can be decomposed into a separable (i.e. unentangled) state by applying the two controlled rotation $\mathbf{R_z}$ and  $\mathbf{R_y}$. By recursively applying this transformation to the $n$-qubit register with the desired target state (i.e. $\ket{G}$, in our case), we can construct a circuit that takes it to the $n$-qubit $ \ket{00...0} $  state. 
  This can be done using  a  quantum multiplexor circuit $ \mathbf{U}$, which is  finally reversed  in order to get the desired initialization circuit. In our case, we identify such an initialization circuit with the unitary $\mathbf{G}$ that manipulates a register of  $N$ qubits initially in $ \ket{0} $ as follows:

\[
 \mathbf{G}\ket{00...0}=\ket{G},
\]
 where  $ \mathbf{G} $ represents the following circuit with elementary components $CX $, $ R_y(\theta) $ and $ R_z(\phi) $:
 \[
 \mathbf{G} =
 \begin{bmatrix}
 R_y(-\theta_0) R_z(-\phi_0) & & & \\
 & R_y(-\theta_1) R_z(-\phi_1) & & \\
 & & \ddots & \\
 & & & R_y(-\theta_{2^{N-1}-1}) R_z(-\phi_{2^{N-1}-1})
 \end{bmatrix}^{{\dagger} }
 \] 
 
 \vspace{0.5cm}
 \noindent
 Unfortunately,  the construction of $ \mathbf{G}$ in this way requires $O(2^{N+1}) $ gates,  thus representing a bottleneck for our algorithm.
It will be the subject of future work to try different state preparation schemes by investigating other approaches such as those in [\citen{Park19,Mottonen05,Arunachalam2015,zhao2018,Park19,ciliberto}].

 \subsection{An Example}
\begin{figure}[htbp]  \vspace{-0.cm}
 	\centering   
 	\includegraphics[width=.9\textwidth]{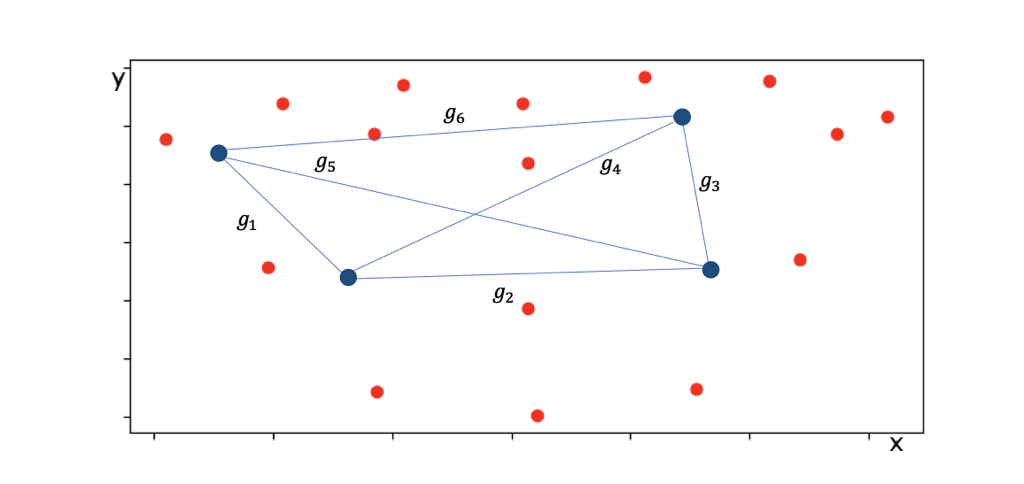}
 	\caption{Complete graph with four randomly selected mouth landmark points.}
 	\label{fig:example}
 \end{figure}

 Let's select  $ n=4 $ random vertices among those belonging to the mouth landmark points. The complete graph constructed for these points looks like the one in Fig.~\ref{fig:example}.

\begin{figure}[htbp]
 		$$
 		\Qcircuit @C=.5em @R=0.5em @!R {
 			& & & & & &\lstick{a: \ket{0}} & \qw & \qw & \qw & \qw& \qw & \qw & \qw & \qw & \qw & \qw & \qw & \gate{\mathbf{X}}& \qw & \ >  \\
 			& & & & & & \lstick{b: \ket{0}} & \qw & \qw & \qw  & \gate{\mathbf{X}} & \qw & \gate{\mathbf{U(0.7854)}}  & \qw  & \gate{\mathbf{X}}  & \qw & \gate{\mathbf{U(0.7854)}}& \qw& \qw & \qw& \ >\\
 			& & & & & & \lstick{c: \ket{0}} & \qw & \gate{\mathbf{U(1.231)}} & \qw & \ctrl{-1}	& \qw & \qw& \qw & \ctrl{-1} & \qw & \qw & \qw & \ctrl{-2} & \qw & \ >	} 
 		$$\\
 			$$
 		\Qcircuit @C=.5em @R=0.5em @!R {
 			& & & & & &\lstick{a: } & \qw &\gate{\mathbf{U(0.3927)}} & \qw &  \gate{\mathbf{X}} & \qw & \gate{\mathbf{U(-0.3927)}} & \qw & \gate{\mathbf{X}} & \qw & \gate{\mathbf{X}} & \qw & \gate{\mathbf{X}} & \qw &\qw &  \ >   \\
 			& & & & & & \lstick{b: } & \qw& \qw & \qw & \ctrl{-1}& \qw& \qw & \qw & \ctrl{-1}& \qw& \qw & \qw & \ctrl{-1}& \qw& \qw  & \ >   \\
 			& & & & & & \lstick{c: } & \qw & \qw& \qw & \qw &\qw &\qw & \qw& \qw & \qw & \ctrl{-2}& \qw& \qw & \qw& \qw& \ >	} 
 		$$
 		\\
 		$$
 		\Qcircuit @C=.5em @R=0.5em @!R {
 			& & & & & &\lstick{a: } & \qw &\gate{\mathbf{U(0.3927)}} & \qw&   \gate{\mathbf{X}} & \qw & \gate{\mathbf{U(1.1781)}} & \qw & \gate{\mathbf{X}} & \qw & \gate{\mathbf{X}} & \qw & \qw  \\
 			& & & & & & \lstick{b: } & \qw& \qw& \qw & \ctrl{-1}& \qw& \qw & \qw & \qw& \qw  & \ctrl{-1}& \qw& \qw  \\
 			& & & & & & \lstick{c: } & \qw& \qw& \qw & \qw& \qw & \qw &\qw & \ctrl{-2} &\qw & \qw& \qw & \qw	} 
 		$$
 		\caption{{Quantum circuit  constructing the state $ \mathbf{G_4}$. Three qubits are employed in the circuit, which are denoted by $a$,  $b$ and $c$.  } }	
		\label{fig:example2}
 	\end{figure}
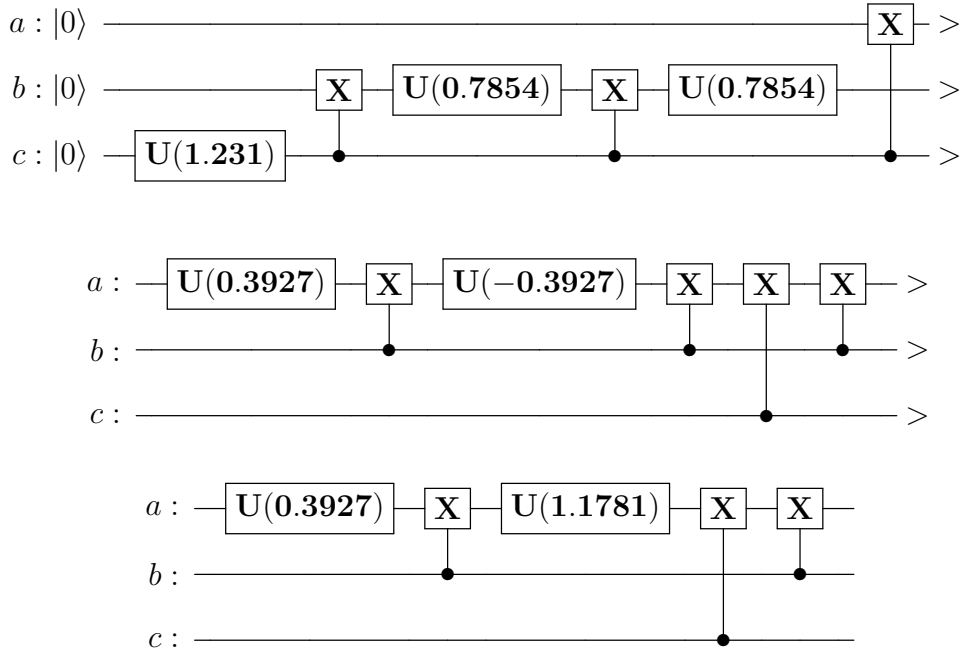

% \FloatBarrier 
 \noindent
  Such a graph is encoded into the quantum state $ \ket{G_4}$ defined by
 \begin{equation*}
 \ket{G_4} = \dfrac{1}{\gamma}\left( g_1 \ket{000}+g_2 \ket{001}+g_3 \ket{010}+g_4 \ket{011}+g_5 \ket{100}+g_6 \ket{101}\right),
 \end{equation*} 
 where $\gamma$ is a normalization constant.
The quantum circuit  $ \mathbf{G_4} $ that realizes  $ \ket{G_4} $, i.e. such that $ \mathbf{G_4}\ket{000}= \ket{G_4}$,  via the method  proposed by Shende et al. in [\citen{Shende}] is shown in Fig.\ref{fig:example2},  where the gate $ \mathbf{U(\theta)} $ is defined by
  \[
   \mathbf{U(\theta)} =
 \begin{bmatrix}
 \cos(\theta/2) & \ \ \ -\sin(\theta/2)   \\
  \sin(\theta/2) &  \ \ \ \cos(\theta/2)  
 \end{bmatrix}
 \]

% \FloatBarrier

 As claimed before, this algorithm is  exponential  in the number of the input qubits and represents a bottleneck to the speed of the whole algorithm. Thus, finding a more efficient encoding of the data into quantum states is crucial for achieving a better performance of the overall algorithm we are going to define in the next section. This would also benefit from a more complex connection between physical qubits in the hardware that would allow us to reduce the number of swap gates used to adapt the quantum  circuit to the topology of the quantum computer.
 %Can you explain this? Is this because with a more complex connection, one may be able to implement multiqubit gates that are more complex than CX?

\section{Graphs Quantum Classifier }
\label{Section3}

The first implementation  of a quantum classifier on the IBM quantum computer appears in [\citen{Fingerhuth}], where data are encoded in a single qubit state. We extend the work in that paper by constructing a circuit that is able to deal with multiple qubit state representations of a dataset.

Given a   dataset $ \mathcal{D} $, consider the  training  set, 
\[
\mathcal{D}_{train} = \{(G^{\left\lbrace 1 \right\rbrace } ,y^{\left\lbrace 1 \right\rbrace }),...,(G^{\left\lbrace M \right\rbrace } ,y^{\left\lbrace M \right\rbrace })\},
\]
of $ M$ pairs  $ (G^{\left\lbrace m \right\rbrace } ,y^{\left\lbrace m \right\rbrace }) $, where   $G^{\left\lbrace m \right\rbrace } $   are graphs (e.g. representing   the face of an individual) while  $y^{\left\lbrace m \right\rbrace }\in \{c_l\}_{l=1} ^L$ identify which of the $ L$  possible  class labels  are  associated to a graph. Classes partition graphs according to some features, which in our case correspond to sad vs happy face expressions.

Given a new unlabeled input $G_{test} $, the task is to assign  a   label ${y_{test}} $ to $G_{test}  $ using     the distance %[\citen{Banks1994}]
\[
d(G_1,G_2)=\sqrt{|\mathbf{a}_{G_{1} }- \mathbf{a}_{G_2 }|^2}= \sqrt{\sum_{i=1}^d \left| g_i^{(G_{1}) }- g_i^{(G_2) }\right| ^2},
\]
and then classifying  according to
\begin{equation}
\min_{c_l} \left[\sum\limits_{m\  s.t.\ y^{\left\lbrace m \right\rbrace }=c_l }  d\left( G_{test} ,G^{\left\lbrace m \right\rbrace }\right)  \right].
\label{Eq:classifier}
\end{equation}

In the  case of binary classification, where there are only two classes, i.e. $y^{\left\lbrace m \right\rbrace }\in 
\{+1,-1\}$,  the classifier of Equation~(\ref{Eq:classifier}) can be expressed as
\begin{equation}
y_{test}=
-\mathrm{sgn} \left[\sum\limits_{m=1}^M y^{\left\lbrace m \right\rbrace } d\left( G_{test} ,G^{\left\lbrace m \right\rbrace }\right)  \right].
\label{eq: model}
\end{equation}

The quantum circuit that implements such a  classification  requires 
four multi-qubit quantum registers. In the initial configuration,
\begin{itemize}
	\item $ \ket{0}_a $ is an ancillary qubit  that is entangled to the qubits encoding both the  test graph and  the training graphs;
	\item $\ket{00..0}_m $ is a register of  $ \lceil \log(M)\rceil$ qubits that stores the index $ m $ of the training graphs $ G^{\left\lbrace m\right\rbrace } $; 
	\item $  \ket{00..0}_g $ is a register of  $ \lceil \log(d)\rceil$ qubits used to encode both test   and training graphs; 
	\item $ \ket{00..0}_c $ is a register of  $ \lceil \log(L)\rceil$ qubits that  stores the classes $ y^{\left\lbrace m \right\rbrace }\in \{c_l\}_{l=1} ^L$ of the $ G^{\left\lbrace m\right\rbrace } $.
\end{itemize}

In the case of a  training set of two graphs, one per class, the circuit is implemented as shown in Fig.~(\ref{Fig:classifier_circuit}).

\begin{center}	
	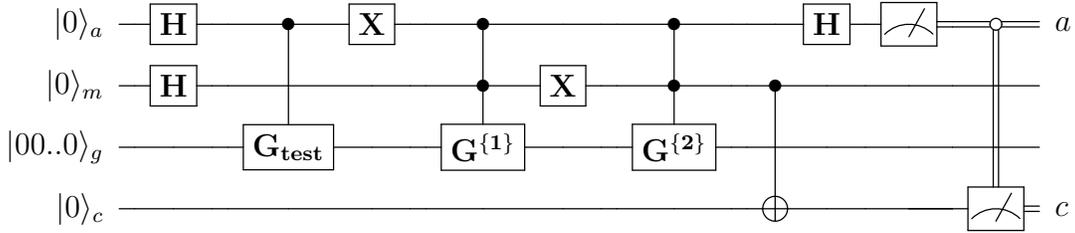
\begin{figure*}[htbp]
		$$
		\Qcircuit @C=.5em @R=0.5em @!R {
			%& & & & & & &  & A &  &  & B &  & &  &  & C &  &  & D &  & &  &  &  & F &  & & \\
			& & & & & &\lstick{\ket{0}_a} & \qw & \gate{\mathbf{H}} & \qw & \qw & \ctrl{2} & \gate{\mathbf{X}} & \qw & \qw & \ctrl{1} & \qw & \qw & \qw & \ctrl{1} & \qw & \qw & \qw & \gate{\mathbf{H}} & \qw & \meter & \cw &  \cctrlo{3} &\rstick{a} \cw \\
			& & & & & & \lstick{\ket{0}_m} & \qw & \gate{\mathbf{H}} & \qw & \qw  & \qw & \qw & \qw & \qw & \ctrl{1} & \gate{\mathbf{X}} & \qw & \qw & \ctrl{1} & \qw & \qw & \ctrl{2} & \qw  & \qw  & \qw & \qw & \qw  & \qw \\
			& & & & & & \lstick{\ket{00..0}_g} & \qw & \qw & \qw & \qw & \gate{\mathbf{G_{test}}}  & \qw & \qw & \qw & \gate{\mathbf{G^{\left\lbrace 1\right\rbrace }}}  & \qw & \qw & \qw & \gate{\mathbf{G^{\left\lbrace 2\right\rbrace }}} & \qw & \ \qw & \qw & \qw  & \qw  & \qw & \qw & \qw &  \qw\\
			& & & & & & \lstick{\ket{0}_c} & \qw & \qw & \qw & \qw  & \qw & \qw & \qw & \qw & \qw & \qw & \qw & \qw & \qw & \qw & \qw  & \targ & \qw & \qw & \qw & \qw\qw & \meter  & \rstick{c} \cw \
		} 
		$$
		\caption{Quantum binary classifier circuit.}\label{Fig:classifier_circuit} 
	\end{figure*}
\end{center}
\noindent
After the first two Hadamard gates, the circuit is in state
\[
\dfrac{1}{2}\left( \ket{0}_a + \ket{1}_a\right) \left( \ket{0}_m + \ket{1}_m\right) \ket{00..0}_g \ket{0}_c.
\]
The $ \mathbf{Control} $-$ \mathbf{G_{test}} $ gate, followed by an $ \mathbf{X} $ operator on the ancilla  produces the state 
\[
\ket{G_{test}} = \dfrac{1}{\gamma_{test}}\sum_{k=1}^{d} g_k^{test} \ket{k},
\]
and entangles it with the $ \ket{0}_a  $  state of the ancilla.
Then a $\mathbf{G^{\left\lbrace 1 \right\rbrace } }$ gate is controlled by both the first qubit $a$ and by the second one $ m $, which is also subjected to an $\mathbf{X}$ gate afterwards. This has the effect of creating the state associated to the training graph $ G^{\left\lbrace 1 \right\rbrace }$, namely
\[
\ket{G^{\left\lbrace 1 \right\rbrace }} = \dfrac{1}{\gamma^{\left\lbrace 1 \right\rbrace }}\sum_{k=1}^{d} g_k^{\left\lbrace 1 \right\rbrace } \ket{k},
\]
and entangling it with both  $ \ket{1}_a  $  and  $ \ket{0}_m $.
Then, the double controlled $  \mathbf{G^{\left\lbrace 2\right\rbrace }} $ gate has a similar  effect, entangling the second training graph state $ \ket{G^{\left\lbrace 2 \right\rbrace }} $ with $\ket{1}_a$  and  $\ket{1}_m$.
At this stage, the $ \mathbf{Control} $-$ \mathbf{X} $ gate entangles the $ m $ index of the two test graphs with the correct class state, i.e. either $ \ket{0}_c$ or $ \ket{1}_c $.
The state of the circuit at this point can be written as 
\[
\frac{1}{\sqrt{2M}} \sum_{m=0}^{M-1}  \Big( \ket{0}_a \ket{G_{test}}_g + \ket{1}_a \ket{G^{\left\lbrace m \right\rbrace }}_g \Big) \ket{m}_m\  \ket{y^{\left\lbrace m \right\rbrace }}_c.
\]
Finally, the Hadamard gate applied to the ancilla generates the state
\[
\frac{1}{2\sqrt{M}} \sum_{m=0}^{M-1}  \Big[\ket{0}_a \left( \ket{G_{test}}_g + \ket{G^{\left\lbrace m \right\rbrace }}_g \right) + \ket{1}_a \left( \ket{G_{test}}_g - \ket{G^{\left\lbrace m \right\rbrace }}_g \right) \Big] \ket{m}_m\  \ket{y^{\left\lbrace m \right\rbrace }}_c.
\]
Measuring the ancilla in the state $ \ket{0}_a  $ produces the state
\[
\frac{1}{\sqrt{ \sum _m \sum _i  |g_i^{(G_{test}) }+ g_i^{(G^{\left\lbrace m \right\rbrace }) }|^2}}  \sum_{m=0}^{M-1}  \sum_{i=1}^d \left( g_i^{(G_{test}) }+ g_i^{(G^{\left\lbrace m \right\rbrace }) }\right)  \ket{i}_g\  \ket{m}_m\  \ket{y^{\left\lbrace m \right\rbrace }}_c,
\]
and a subsequent measurement of the class qubit $ \ket{y^{\left\lbrace m\right\rbrace}} $ gives the outcome corresponding to state $ \ket{0}_c $ with probability
\begin{equation*}
\begin{split}
\mathrm{p}(y^{\left\lbrace m\right\rbrace} = 0)= \frac{1}{{ \sum _m \sum _i |g_i^{(G_{test}) }+ g_i^{(G^{\left\lbrace m \right\rbrace }) }|^2}} \sum\limits_{m\ s.t.\  y^{\left\lbrace m\right\rbrace}  = 0}  \  \sum_{i=1}^d \left| g_i^{(G_{test}) }+ g_i^{(G^{\left\lbrace m \right\rbrace }) }\right|^2=\\= 1-\left( \frac{1}{{ \sum _m \sum _i |g_i^{(G_{test}) }+ g_i^{(G^{\left\lbrace m \right\rbrace }) }|^2}} \sum\limits_{m\ s.t.\  y^{\left\lbrace m\right\rbrace}  = 0}  \  d(G_{test}, G^{\left\lbrace m \right\rbrace })^2\right).
\end{split}
\end{equation*}

\noindent
This probability depends on the distance between  the test graph $ G_{test} $ and all those training graphs belonging  to the class $ c=0 $.
Therefore, if its value is  lower than $ 0.5 $, then the test set is classified as  $ y_{class}=-1 $ while if it is higher, then the test set is classified as  $y_{class}=+1 $.
The quantum circuit hence realizes the classification expressed by  Equation~(\ref{eq: model}).

\section{Classification Algorithm and Experimental Results }
\label{algos}
The number of qubits used by the quantum classifier scales with  the number, $ n $, of vertices in the graphs, which are randomly selected  among those of the mouth landmark points.  As shown in Table~\ref{tab:input-size} the encoding allows for  an exponential compression of resources  since only a number $ O(log_2(n^2)) $ of qubits are  needed to represent a complete graph of $ n $ nodes.
In the table, the number of elements in the adjacency matrix that identify the graphs is denoted by $ d $ and its value is $n(n-1)/2$. 

\begin{table}[htbp]
	\begin{center}
		\begin{tabular}{ |c |c |c |}
			\hline
			$ n$    & $ d \approx O(n^2) $  & $\#	$  qubits used by the algorithm  \\ \hline
			3 &  9 & 7 \\ \hline
			4 &  16 & 8 \\ \hline
			5 &  25 & 8 \\ \hline
			10&  100 & 10 \\ \hline
			20 &  400 & 12 \\ \hline
			50&  2500 & 15 \\ \hline
			100 &  10000 & 17 \\ \hline
			500	& 250000	& 21 \\ \hline
			1$\times10^{7}$	& 1$\times10^{14}$	& 50 \\ \hline		
		\end{tabular}	
		\caption{\label{tab:input-size}Input size of the classification circuit depending on the number of vertices.}
	\end{center}
\end{table}
\FloatBarrier

We evaluated our quantum classifier on  both the complete and the meshed graphs using the \texttt{qasm\_simulator} available through the IBM cloud platform.
We compared the results with a classical binary classifier based on the 
Fr\"obenius norm between graph adjacency matrices  defined by
\[
||A|| = \sqrt{\sum_{i, j} | a_{ij} |^2},
\]
so that  the distance used in Equation~(\ref{eq: model}) becomes
\begin{equation}\label{eq:distance}
d\left( G_{test}, G^{\left\lbrace m \right\rbrace }\right) =\sqrt{\sum_{ij} {| (g_i^{test} - g_i^{\{m\}})|}^2}.
\end{equation}

In order to perform our experiments, we proceeded as follows.
%\begin{itemize}
%	\item 
	By randomly selecting 10 different test graphs $ G_\test$ and  25 different pairs of labeled graphs 
	$\left\lbrace  G_\sad, G_\happy \right\rbrace$, we constructed 250  items of the form $ \left\lbrace G_\test, G_\sad, G_\happy \right\rbrace  $ for the classical classifier and the   IBM  quantum simulator. %\texttt{qasm\_simulator}.
%\end{itemize}
Vertices were selected randomly among the 20 landmark points of the mouth, the number $ n $ of vertices given in input to the algorithm  was gradually increased, and a  one-time sample of the vertices was shared with all methods. 

\noindent
Based on this, we have devised an algorithm for classification that is divided into two main steps.  
The first step consists in a procedure, named \texttt{ClassifyWrtSingleFace}, that considers each item  $\left\lbrace G_\test, G_\sad, G_\happy \right\rbrace $ individually and  classifies $G_{test}$  as happy or sad, based only on the  distances 
$d(G_\test, G_\sad)$ and  $d(G_\test, G_\happy) $. The procedure  \texttt{AccuracyWrtSingleFace} evaluates the accuracy of such a classification by essentially counting  the number of correct answers for all the test graphs.
In the second step, another procedure, named \texttt{ClassifyWrtWholeSet}, performs a different classification where each  face $ G_\test $ is labeled depending on the output of \texttt{ClassifyWrtSingleFace} calculated on the whole set  $\{G_\happy, G_\sad \} $.

We  have compared the accuracy obtained with different number $n$ of landmark points of the mouth ranging from $4$ to $20$. 
The $20$ points case considers all the points of the mouth. 
The $n < 20$ cases consider  subsets of the mouth landmark points  which  are chosen randomly and uniformly in the coordinates. For each $n$ we have picked up 20  randomly chosen subsets, and we have calculated the accuracy of the classifier for each $n$  as the mean of the accuracies obtained by varying such subsets.

\subsection{Algorithm Step 1}
\begin{algorithmic}[1]
	\Procedure{ClassifyWrtSingleFace}{$G_\test, G_\sad, G_\happy$}
	\State $d \gets \fun{Distance}(G_\test, G_\sad) - \fun{Distance}(G_\test, G_\happy)$
	\If{abs($d$) < 0.001}
	\State \return unknown
	\EndIf
	\If{$d$ < 0}
	\State \return \happy
	\Else
	\State \return \sad
	\EndIf
	\EndProcedure
\end{algorithmic}
\noindent \newline
The evaluation of the distance is obtained classically via the calculation of  Equation~(\ref{eq:distance}), and quantumly via the application of the   circuit to the item  $ \{G_\test, G_\sad, G_\happy\} $. The circuit is executed  1024 times.

Given  the set of all test graphs $\mathcal{G} = \{ (G_i, \ell_i) \mid i = 1, ..., n \}$, the following procedure describes the calculation of the accuracy of the classification with respect to a single face.
\\
\begin{algorithmic}[1]
	\Procedure{CalculateAccuracyWrtSingleFace}{$\G$}
	\State correct $\gets 0$
	\State wrong $\gets 0$
	\State $\G_\sad = \{ G \mid (G, \ell) \in \G \land \ell = \sad \}$
	\State $\G_\happy = \{ G \mid (G, \ell) \in \G \land \ell = \happy \}$
	\ForAll{$(G_\test, \ell_\test) \in \G$}
	\ForAll{$G_\sad \in \G_\sad$, $G_\happy \in \G_\happy$}
	\If{$\fun{ClassifyWrtSingleFace}(G_\test, G_\sad, G_\happy) = \ell_\test$}
	\State correct $\gets$ correct $ + 1$
	\Else
	\State wrong $\gets$ wrong $ + 1$
	\EndIf
	\EndFor
	\EndFor
	\State AccuracyWrtSingleFace $\gets$ correct$/($correct$+$wrong$)$
	\State \return AccuracyWrtSingleFace
	\EndProcedure
\end{algorithmic}
\noindent\newline

\begin{figure}[htbp] 
    \centering
	\includegraphics[width=.8\textwidth]{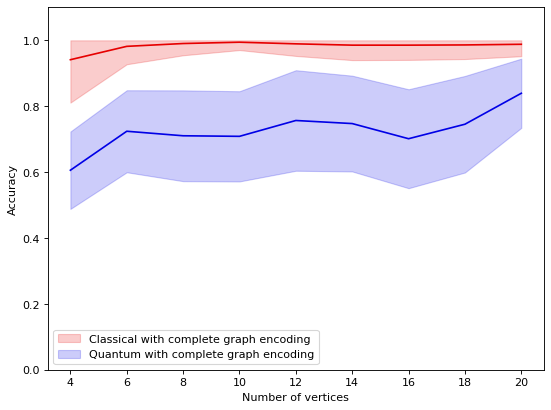}
	
	\includegraphics[width=.8\textwidth]{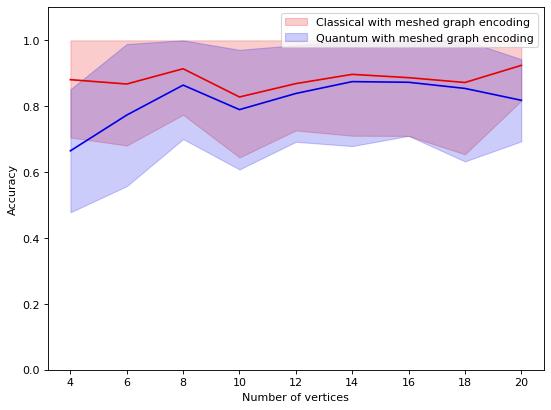}
	
	\caption{Results of the procedure \texttt{AccuracyWrtSingleFace} on graphs with number of vertices $n\in [4,20]$, obtained with the \texttt{qasm\_simulator} compared with the classical classifier. Plots show the average accuracy of the classifiers with  shadow indicating the max and min accuracies.}
\label{fig:ACCURACYWRTSINGLEFACE}
\end{figure}

The values of the accuracy of procedure \texttt{ClassifyWrtSingleFace} that we obtained  with the \texttt{qasm\_simulator}  are reported in Table~\ref{Tacc} and described by the plot in Fig.~\ref{fig:ACCURACYWRTSINGLEFACE}.

%\FloatBarrier

\begin{table}[htbp]
	\begin{center}
		\begin{tabular}{ |c |c |c |c |c |}
			\hline
			$ n$    & CC meshed  & CC complete & QC meshed & QC complete \\ \hline	
				4   & 0.88  & 0.94 & 0.66 & 0.61 \\ \hline	
				6   & 0.87 & 0.98 & 0.77 & 0.72\\ \hline	
				8   & 0.91  & 0.99 & 0.86& 0.71 \\ \hline	
				10  & 0.83  & 0.99 & 0.79 & 0.71\\ \hline
				12  & 0.87 & 0.99 & 0.84 & 0.76 \\ \hline		
				14  & 0.90  & 0.98 & 0.87& 0.75 \\ \hline	
				16  & 0.89  & 0.98 & 0.87 & 0.70\\ \hline						
				18  & 0.87  & 0.99 & 0.85 & 0.75\\ \hline	
				20  & 0.92  & 0.99 & 0.82& 0.84 \\ \hline	
		\end{tabular}	
	\end{center}
	\caption{\label{Tacc} Average accuracy of the \texttt{ClassifyWrtSingleFace} procedure obtained with the \texttt{qasm\_simulator} (QC), and the classical  classifier (CC).}
\end{table}
%\FloatBarrier

Fig.~\ref{fig:ACCURACYWRTSINGLEFACE} and Table~\ref{Tacc} show that the highest value of the \texttt{AccuracyWrtSingleFace} procedure is  obtained with the classical classifier, using the complete graph strategy. The classical classifier with meshed graphs reveals a worse performance. This is due to the fact that  meshed graphs, on  average, carry less information about the shape of the mouth.  The performance gap between the complete and meshed approach is not so evident for what concerns the quantum simulation. 
This may be due to the fact  that the encoding of a complete graph into a quantum state is much more complex than for meshed graphs since for complete graphs there are always $ n(n-1)/2 $ amplitudes that are different from zero. In the meshed case, the graphs are sparse and much less amplitudes are needed  in the quantum states for encoding them. Clearly, the complexity of the quantum states negatively affects the robustness of the circuit. Moreover, it is interesting to notice how the accuracy of the quantum simulation with meshed graphs closely follows the trend of the accuracy obtained  by its classical counterpart.

%The quantum computation only allowed us to test graphs with up to 8 vertices; after that we couldn't get any outcome from the quantum computer, probably because the number of gates necessary for $n > 8$ exceeded the coherence time of the qubits. Moreover, the results for the accuracy that we obtained on the quantum computer are biased by the fact that the classification has been performed only on  16 items  of the kind $ \{G_\test, G_\sad, G_\happy\}$ compared to the 250 items that we used for the classical  case and the quantum simulation.

\subsection{Algorithm Step 2}
We now refine the classification by comparing $G_\test$ with the entire set  of graphs, $ \{G_\happy, G_\sad \}$, so that a face is classified as  $\happy$ if by running procedure \texttt{ClassifyWrtSingleFace} we obtain more times  $\happy$ than $\sad$, and $\sad$ otherwise. 

The \texttt{ClassifyWrtWholeSet} procedure is described below.
\\\\
\begin{algorithmic}[1]
	\Procedure{ClassifyWrtWholeSet}{$G_\test, \G$}
	\State happycounter $\gets 0$
	\State sadcounter $\gets 0$
	\State $\G_\sad = \{ G \mid (G, \ell) \in \G \land \ell = \sad \}$
	\State $\G_\happy = \{ G \mid (G, \ell) \in \G \land \ell = \happy \}$
	\ForAll{$G_\sad \in \G_\sad$, $G_\happy \in \G_\happy$}
	\If{$\fun{ClassifyWrtSingleFace}(G_\test, G_\sad, G_\happy) = \happy$}
	\State happycounter $\gets$ happycounter $ + 1$
	\Else
	\State sadcounter $\gets$ sadcounter $ + 1$
	\EndIf
	\EndFor
	\If{happycounter$==$sadcounter}
	\State \return unknown
	\EndIf
	\If{happycounter$>$sadcounter}
	\State \return \happy
	\Else
	\State \return \sad
	\EndIf
	\EndProcedure
\end{algorithmic}
\noindent\newline

The accuracy of the above procedure is calculated by the following algorithm as the frequency of the correct answers returned by the \texttt{ClassifyWrtWholeSet} procedure. 
\\\\
\begin{algorithmic}[1]
	\Procedure{CalculateAccuracyWholeSet}{$\G$}
	\State correct $\gets 0$
	\State wrong $\gets 0$
	\ForAll{$(G_\test, \ell_\test) \in \G$}
	\If{$\fun{ClassifyWrtWholeSet}(G_\test, \G) = \ell_\test$}
	\State correct $\gets$ correct $ + 1$
	\Else
	\State wrong $\gets$ wrong $ + 1$
	\EndIf
	\EndFor
	\State AccuracyWholeSet $\gets$ correct$/($correct$+$wrong$)$
	\State \return AccuracyWholeSet
	\EndProcedure
\end{algorithmic}
\noindent\newline

\begin{figure}[htbp] 
    \centering
	\includegraphics[width=.8\textwidth]{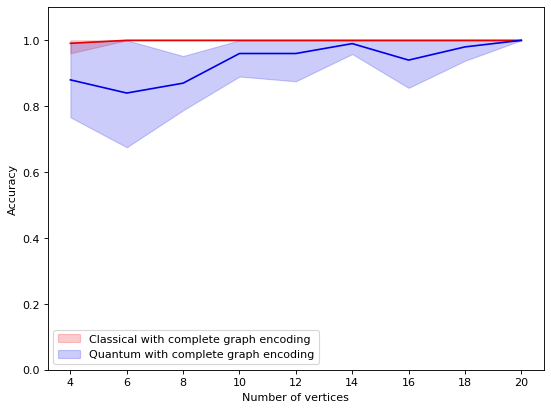}
	
	\includegraphics[width=.8\textwidth]{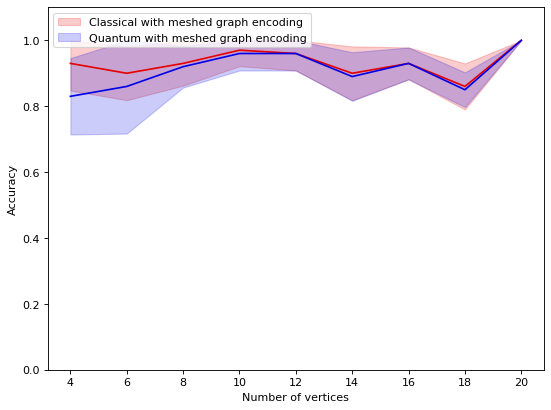}
	
	\caption{ Results of the procedure \texttt{AccuracyWholeSet} on graphs with number of vertices $n\in [4,20]$, obtained with the \texttt{qasm\_simulator}  compared with the classical classifier. Plots show the average accuracy of the classifiers with  shadow indicating the max and min accuracies.}
    \label{fig:accuracyreal}
\end{figure}

The values returned by the  \texttt{CalculateAccuracyWholeSet} procedure for the \texttt{qasm\_simulator} and   the classical binary classifier  are shown in Fig.~\ref{fig:accuracyreal} and Table~\ref{WTacc}.
\\

%\FloatBarrier
%\noindent 

\begin{table}[htbp]
	\begin{center}
		\begin{tabular}{ |c |c |c |c |c |c |c |}
			\hline
			$n$    & CC meshed & CC complete & QC meshed & QC complete  \\ \hline	
			4   & 0.93 & 0.99 & 0.83  & 0.88 \\ \hline	
			6   & 0.90 & 1.0 & 0.86  & 0.84 \\ \hline	
			8   & 0.93 & 1.0 & 0.92  & 0.87 \\ \hline	
			10  & 0.97 & 1.0 & 0.96  & 0.96 \\ \hline
			12  & 0.96 & 1.0 & 0.96  & 0.96 \\ \hline		
			14  & 0.90 & 1.0 & 0.89  & 0.99 \\ \hline	
			16  & 0.93 & 1.0 & 0.93  & 0.94 \\ \hline						
			18  & 0.86 & 1.0 & 0.85  & 0.98 \\ \hline	
			20  & 1.0 & 1.0 & 1.0  & 1.0 \\ \hline	
		\end{tabular}	
		\caption{\label{WTacc}Results of the average \texttt{AccuracyWholeSet} procedure on graphs with number of vertices $n\in [4, 20]$, obtained with the \texttt{qasm\_simulator} (QC) and the classical  classifier (CC).}
	\end{center}
\end{table}
%\FloatBarrier
As in Step 1, the highest value of the accuracy is  obtained with the classical classifier using the complete graph strategy: the fact that  all distances between landmark points are considered makes the classifier very faithful. However, it is worth noticing that we only considered small graphs (up to 20 nodes), thus reducing the amount  of resources  needed to deal with large complete graphs. 
In  applications  where graph dimensions  are several orders of magnitude higher,  a complete graph classification approach may not be a viable option.

A trade-off between graph complexity and accuracy of the classification is reached  in the classical case by employing the meshed graphs encoding;  good  levels of accuracy are obtained with our \texttt{AccuracyWholeSet} algorithm, which never go below  $0.86$.
%As explained before, the alternating values of the accuracy are due to the fact that  meshed graphs may  vary substantially by increasing the number of nodes. Nevertheless, we observe that at $n=10$ and $n=20$ the meshed graph accuracy matches the one obtained with complete graphs.

In the quantum simulation, the complete graph strategy performs clearly worse than its classical counterpart, while the quantum meshed approach shows an average accuracy which is  very close to the classical one.

Comparing Fig.~\ref{fig:accuracyreal} and Fig.~\ref{fig:ACCURACYWRTSINGLEFACE}, we  can also observe that in the latter, the quantum complete graph strategy has a smaller deviation (blue shadows in Fig.\ref{fig:ACCURACYWRTSINGLEFACE} ) than for the quantum meshed graph approach; this is not the case for  \texttt{AccuracyWholeSet}, where we achieve comparable results.
%\textcolor{red}{This is due to the fact that the quantum complete graph  strategy has smaller errors (shadows in Fig.\ref{fig:ACCURACYWRTSINGLEFACE} ) for the \texttt{AccuracyWrtSingleFace} than for the quantum meshed graph approach.

%Moreover, it is important to notice  that the quantum simulation of the meshed method produces an accuracy that is close and sometime above the accuracy of the classical counterpart. 

%The real quantum computer only allowed for the classification of graphs with up to 8 vertices; we couldn't get any outcome for bigger graphs.  Possible explanations for this are the limited access we had to the device and a number of gates  in the circuit exceeding the coherence time of the qubits. Moreover, results of the accuracy obtained on the quantum computer are biased by the fact that the classification has been performed only on 16 items of the kind $ \{G_\test, G_\sad, G_\happy\} $.
%Dan_comment: Can you provide a rough analysis on the hardware requirement (number of qubits, gate fidelity, etc.) for solving an interesting, beyond proof-of-principle, classification problem?

\section{Conclusion}
\label{Section4}

In this work  we have addressed the problem of classifying graphs representing human facial expressions.  
Graphs are obtained by selecting some of the face landmark points and connecting them either via a triangulation of the points or via a complete graph construction.
We have shown the construction of quantum states whose amplitudes encode the graphs, and  devised an interference-based quantum circuit  that performs  a distance-based  classification for the encoded graphs. 
We have implemented this quantum circuit by means of the IBM open-source quantum computing framework Qiskit [\citen{Qiskit}], and tested it on a collection of images taken from the Cohn-Kanade (CK) database, one of the most widely used test-beds for algorithm development and evaluation in the field of automatic facial expressions recognition. 

The tests have been performed on the \texttt{qasm\_simulator}, and the results have been  compared with the classical classifier.
A calculation of the accuracy revealed  that the classical classifier achieves better results when a complete graph approach is employed, while the simulated quantum classifier  achieves comparable results with both the meshed and the complete strategy.

Future work should investigate the possibilities of different and more efficient graph encodings into quantum states.  Moreover, interesting extensions of the results we have presented could be obtained  by exploiting other quantum  classification schemes, such as those based on  the SWAP test along the lines of [\citen{Park_2020}], or on Variational Quantum Circuits [\citen{McClean_2016}].

%\section{Acknowledgements}
%We acknowledge the use of the IBM Q for this work. The views expressed are those of the authors and do not reflect the official policy or position of IBM or the IBM Q team.

\bibliographystyle{spmpsci.bst}      % mathematics and physical sciences
\bibliography{qml}   % name your BibTeX data base

%------------------------------------------

\end{document}